\documentclass[a4paper,12pt]{elsarticle}

\usepackage[ruled,vlined]{algorithm2e}
\usepackage{textcomp}
\usepackage{adjustbox}
\usepackage{mathtools}
\usepackage{booktabs}
\usepackage{etoolbox}
\usepackage[group-separator={,}]{siunitx}
\usepackage{changepage}
        
\usepackage{pgfplots}    
\usepackage{color}
\usepackage{url}

\newcommand{\blue}[1]{{\color{blue}#1}}
\newcommand{\green}[1]{{\color{green}#1}}
\renewcommand{\blue}[1]{#1}
\renewcommand{\green}[1]{#1}

\usepackage{graphics}
\usepackage{latexsym}
\usepackage{rotating}
\usepackage[utf8]{inputenc}

\usepackage{amsmath, amssymb}
\usepackage{index}
\usepackage{stmaryrd}
\usepackage{leftidx}
\usepackage{mathabx}
\usepackage{tikz}
\usepackage{longtable}
\usepackage{hyperref}
\usetikzlibrary{arrows,matrix,positioning,calc}

\newindex{default}{idx}{ind}{Index}

\newcommand{\noticka}[1]{}
\def\tu#1{\langle #1\rangle}

\def\Impl{\!\Rightarrow\!}

\newcommand{\Int}{\ensuremath{\mathrm{Int}}}

\newcommand{\up}{{\uparrow}}
\newcommand{\down}{{\downarrow}}

\newcounter{dct}

\newtheorem{theorem}{Theorem}

\begin{document}
\begin{frontmatter}
\title{LinCbO: fast algorithm for computation of the Duquenne-Guigues basis}

\author{Radek Janostik}
\ead{radek.janostik@upol.cz}

\cortext[cor1]{Corresponding author}

\author{Jan Konecny\corref{cor1}}
\ead{jan.konecny@upol.cz}

\author{Petr Krajča}
\ead{petr.krajca@upol.cz}

\address{Dept.\,Computer Science, Palack\'y University Olomouc \\
	17.\,listopadu 12, CZ--77146 Olomouc\\
	Czech Republic}

\begin{abstract}
We propose and evaluate a novel algorithm for computation of the Duquenne-Guigues basis which combines Close-by-One and LinClosure algorithms.
This combination enables us to reuse attribute counters used in LinClosure and speed up the computation.
Our experimental evaluation shows that it is the most efficient algorithm for computation of the Duquenne-Guigues basis.
\end{abstract}
\end{frontmatter}

\section{Introduction}
\label{sec:intro}

Formal Concept Analysis \cite{GaWi:FCA,Carpineto:2004aa} (FCA)
has two main outputs: (i) hierarchy of formal concepts, called a concept lattice,
in the input data
and (ii) 
a non-redundant system of attribute implications, called a basis, describing the input data.
For both of these outputs, closure systems are the fundamental structures
behind the related theory and algorithms.

Many algorithms for computing closure systems exist \cite{Carpineto:2004aa,serser}.
Among the most efficient algorithms 
are variants of Kuznetsov's Close-by-One (CbO) \cite{kuznetsov1993fast},
namely Outrata \& Vychodil's FCbO \cite{Outrata2012114} and Andrews's In-Close
family of algorithms \cite{Andrews2009,Andrews:2011:IHP:2032828.2032834,Andrews2015633,Andrews2017,andrews2018new}.
These are commonly used for enumeration of formal concepts, as both their parts, extents and intents, form a closure systems.

When considering systems of attribute implications, 
pseudo-intents play an important role, since they derive 
the minimal basis, called the Duquenne-Guigues basis or canonical basis \cite{GuDu:Fmiirtdb}.
The pseudo-intents, together with the intents of formal concepts, form 
a closure system.
Enumerating all pseudo-intents (together with intents) 
is more challenging 
as it requires a particular restriction of the order of the computation
and
the results on complexity are all but promising \cite{Kuznetsov04intractability}. 
There are basically two main approaches for this task: 
NextClosure by Ganter \cite{Ganter1991,GaWi:FCA}, and
the incremental approach by Obiedkov and Duquenne \cite{obiedkov2007attribute}.%



We present a new approach based on the CbO algorithm and LinClosure \cite{maier1983theory}.\footnote{LinClosure is an algorithm for computation of the smallest model of a theory containing a given set of attributes. It uses so-called attribute counters to avoid set comparisons and reach a linear time complexity. We recall this in Section~\ref{sec:linclosure}.} 
\green{Putting it simply, we enumerate members of the closure system (intents and pseudo-intents)
using CbO while each member is computed using LinClosure.}
We show that in our approach,
LinClosure is able to reuse attribute counters from previous computations.
This makes it work very fast, as our experiments show.

The rest of the paper has the following structure: 
First, we recall basic notions of FCA (Section~\ref{sec:pre:fca}),
closure operators (Section~\ref{sec:pre:clo}), 
bases of attribute implications (Section~\ref{sec:pre:bases}),
the algorithm CbO \blue{and NextClosure}
(Section~\ref{sec:cbo}), and
the \blue{algorithms LinClosure (Section~\ref{sec:linclosure})
and Wild's closure (Section~\ref{sec:wildclosure})
}. Second, we introduce our approach, which includes
CbO with changed sweep order (Section \ref{sec:sweep_order}) and improvements previously introduced into NextClosure in \cite{bazhanov2014} (Section \ref{sec:nc_improv}). Most importantly, we describe a feature
which enables LinClosure to reuse the attribute counters (Section \ref{sec:imlc}). Then, we experimentally evaluate the resulting algorithm
(Section~\ref{sec:ee}) and discuss our observations (Section~\ref{sec:discus}).
Finally, we summarize our conclusions and present ideas for further research (Section~\ref{sec:conc}).

\section{Preliminaries}

Here, we recall notions used in the rest of the paper.

\subsection{Formal concept analysis}
\label{sec:pre:fca}

An input to FCA is a triplet $\tu{X,Y,I}$,
called a \emph{formal context}, where $X,Y$ are non-empty sets of objects and attributes respectively, and $I$ is a binary relation between $X$ and $Y$.
The presence of an object-attribute pair $\tu{x,y}$ in 
the relation $I$ means that the object $x$ has the attribute $y$.

Finite contexts are usually depicted as tables, in which rows represent objects in $X$, columns represent attributes in $Y$, ones in its entries mean that the corresponding object-attribute pair is in $I$.


The formal context $\tu{X,Y,I}$ induces so-called {\it concept-forming operators}:
\begin{itemize}
\item[]
${}^\up: \mathbf{2}^X \to \mathbf{2}^Y$ assigns to a set $A$ of objects the set $A^\up$ of all attributes shared by all the objects in $A$.
\item[]
${}^\down:\mathbf{2}^Y \to \mathbf{2}^X$ assigns to a set $B$ of attributes the set $B^\down$ of all objects which share all the attributes in $B$.
\end{itemize}
Formally, for all $A\subseteq X, B \subseteq Y$ we have
\begin{align*}
A^\up &= \{y \in Y\mid \forall x \in A\,:\,\tu{x,y} \in I\}, \\
B^\down &= \{x \in X\mid \forall y \in B\,:\,\tu{x,y} \in I\}.
\end{align*}

Fixed points of the concept-forming operators, i.e. pairs $\tu{A,B} \in \mathbf{2}^X \times \mathbf{2}^Y$ satisfying 
$A^\up = B$ and  $B^\down = A$,
are called {\em formal concepts}. 
The sets $A$ and $B$ in a formal concept $\tu{A,B}$
are called the {\em extent} and the {\em intent}, respectively.

The set of all intents in $\tu{X,Y,I}$ is denoted by $\Int(X,Y,I)$.

\medskip

An {\em attribute implication} is an expression of the form $L \Impl R$
where $L,R \subseteq Y$ are sets of attributes.

We say that $L \Impl R$ is valid in a set of attributes $M \subseteq Y$ if
\[
 L \subseteq M \text{ implies } R \subseteq M.
\]
The fact that $L \Impl R$ is valid in $M$ is written as $\|L \Impl R\|_M=1$.

We say that $L \Impl R$ is valid in a context $\tu{X,Y,I}$ if
it is valid in every object intent $\{x\}^\up$, i.e.
\[
\|L \Impl R\|_{\{x\}^\up}=1 \qquad \forall x\in X.
\]
A set of attribute implications is called a {\em theory}.

\newcommand{\Mod}{{\mathrm{Mod}}}

A set of attributes $M$ is called a {\em model} of theory $\cal T$ if every attribute
implication in $\cal T$ is valid in $M$. The set of all models of $\cal T$ is
denoted $\Mod({\cal T})$, i.e.
\[
\Mod({\cal T}) = \{ M \mid  \forall L \Impl R \in {\cal T} :  \| L \Impl R \|_M = 1 \}.
\]

\subsection{Closure systems and closure operators}
\label{sec:pre:clo}

A {\em closure system} in a set $Y$ is any system $\cal S$ of subsets of $Y$ which
contains $Y$ and is closed under arbitrary intersections.

A {\em closure operator} on a set $Y$ is a mapping $c: 2^Y \to 2^Y$ satisfying for each
$A,A_1,A_2 \subseteq Y$:
\begin{align}
&A \subseteq c(A)\\
\label{eq:co:monotony}
&A_1 \subseteq A_2 \text{~~~implies~~~} c(A_1) \subseteq c(A_2) \\
&c(A) = c(c(A)).
\end{align}
The closure systems and closure operators are in one-to-one correspondence.
Specifically, for a closure system $\cal S$ in $Y$, the mapping 
$c_{\cal S}: 2^Y \to 2^Y$ defined by
\[
c_{\cal S}(A) = \bigcap\{ B \in {\cal S} \mid A \subseteq B \}
\]
is a closure operator. Conversely, for a closure operator $c$ on $Y$, the set
\[
{\cal S}_c = \{ A \in 2^Y \mid c(A) = A  \}
\]
is a closure system. Furthermore, ${\cal S}_{c_{\cal S}} = {\cal S}$ 
and ${c_{{\cal S}_c}} = c$.
\medskip

For a formal context $\tu{X,Y,I}$, the set $\Int(X,Y,I)$ of its intents 
is a closure system. The corresponding closure operator, $c_{\Int(X,Y,I)}$,
is equal to the composition $^{\down\up}$ of concept-forming operators.

For any theory $\cal T$, the set $\Mod({\cal T})$ of its models is 
a closure system. The corresponding closure operator, $c_{\Mod({\cal T})}$,
is equal to the following operator $c_{\cal T}$.
For $Z \subseteq Y$ and theory $\cal T$, put
\begin{enumerate}
\item $Z^{\cal T} = Z \cup \bigcup \{ R \mid L \Impl R \in T, L \subseteq Z \},$
\item $Z^{{\cal T}_0} = Z,$
\item $Z^{{\cal T}_n} = (Z^{{\cal T}_{n-1}})^{\cal T}.$
\end{enumerate}
Define operator $c_{\cal T}: 2^Y \to 2^Y$ by 
\[
c_{\cal T}(Z) = \bigcup_{n=0}^\infty Z^{{\cal T}_n}.
\]

\subsection{Bases, Duquenne-Guigues basis and its computation}
\label{sec:pre:bases}
A theory $\cal T$ is called 
\begin{itemize}
\item
{\em complete} in $\tu{X,Y,I}$ if $\Mod({\cal T}) = \Int(X,Y,I)$;
\item
a {\em basis} of $\tu{X,Y,I}$ if no proper subset of ${\cal T}$ is 
complete in $\tu{X,Y,I}$.
\end{itemize}

A set $P \subseteq Y$ of attributes is called a {\em pseudo-intent} if it satisfies
the following conditions:
\begin{itemize}
\item[(i)] it is not an intent, i.e. $P^{\down\up} \neq P$; 
\item[(ii)] for all smaller pseudo-intents $P_0 \subset P$, we have $P_0^{\down\up} \subset P$.
\end{itemize}

\begin{theorem}
\label{thm:gd}
Let $\cal P$ be a set of all pseudo-intents of $\tu{X,Y,I}$. The set 
\[
\{
P \Impl P^{\down\up} \mid P \in {\cal P}
\}
\]
is a basis of $\tu{X,Y,I}$. Additionally, it is a minimal basis in terms of the number of attribute implications.
\end{theorem}

The basis from Theorem~\ref{thm:gd} is called the {\em Duquenne-Guigues basis}.

Let $\cal P$ be a set of all pseudo-intents of $\tu{X,Y,I}$.
The union $\Int(X,Y,I) \cup {\cal P}$ is a closure system on $Y$.

The corresponding closure operator $\tilde{c}_{\cal T}$ is given as follows.
For $Z \subseteq Y$ and theory $\cal T$, put
\begin{enumerate}
\item $Z^{\cal T} = Z \cup \bigcup \{ R \mid L \Impl R \in {\cal T}, L \subset Z \},$
\item $Z^{{\cal T}_0} = Z,$
\item $Z^{{\cal T}_n} = (Z^{{\cal T}_{n-1}})^{\cal T}.$
\end{enumerate}
Define operator $\tilde{c}_{\cal T}: 2^Y \to 2^Y$ by 
\begin{align}
\label{eq:closure_operator}
\tilde{c}_{\cal T}(Z) = \bigcup_{n=0}^\infty Z^{{\cal T}_n}.
\end{align}

The algorithm which follows the above definition is called the na\"ive algorithm.
There are more sophisticated ways to compute closures,
like
LinClosure \cite{maier1983theory},
Wild's closure \cite{wild1995computations}\blue{,
and $\mathbf{SL}$-closure \cite{slclosure}.}

Note that the definition of $\tilde{c}_{\cal T}$ differs from the definition of $c_{\cal T}$ in Section~\ref{sec:pre:clo}
only in the subsethood in item 1 -- the operator $c_{\cal T}$ allows equality in this item while $\tilde{c}_{\cal T}$ 
does not. 
In what follows, we use the shortcut $Z^\bullet$ for $\tilde{c}_{\cal T}(Z)$.

\blue{
Let $Z$ be a set of attributes and ${\cal S}$ be a subset of attribute implications such that
\begin{equation}
\label{eq:huhuhu}
\begin{minipage}{.8\textwidth}
\begin{itemize}
\item all implications $L \Impl R \in {\cal T}$ with $L \subset Z^\bullet$ are in $\cal S$,
\item no attribute implication $L \Impl R \in {\cal T}$ with $L=Z^\bullet$ is in $\cal S$.
\end{itemize}
\end{minipage}
\end{equation}
Then, we clearly have, $c_{\cal S}(Z) = Z^\bullet$.

This gives a basic picture, how we compute the Duquenne-Guigues basis $\cal T$:
starting with ${\cal S} =\emptyset$, we compute $c_{\cal S}(Z)$ for a set $Z$ for which $\cal S$
satisfies the conditions \eqref{eq:huhuhu}. 
If $Z^\bullet$ is a pseudo-intent, we update ${\cal S}$ by adding the attribute implication
$Z^\bullet\Impl Z^{\bullet\down\up}$, and repeat for other sets $Z$. When all plausible 
sets are processed, $\cal S$ is the Duquenne-Guigues basis $\cal T$.

Therefore,}
the intents and pseudo-intents must be enumerated in an order $\le$
which extends the subsethood; i.e. 
\begin{equation}
\label{eq:subsethood}
C_1 \subseteq C_2\text{ implies }C_1 \le C_2\qquad\text{for all }C_1,C_2 \in \Int(X,Y,I) \cup {\cal P}.
\end{equation}
\blue{
NextClosure enumerates closed sets in so-called {\em lectic order}.
We obtain
the lectic order of sets
      when we order 
      their
      characteristic vectors
      as binary numbers.      
      }
The lectic order 
satisfies \eqref{eq:subsethood}; that is why NextClosure \cite{GaWi:FCA} \blue{(described at the end of Section~\ref{sec:cbo})} is most frequently used for \blue{the computation of the Duquenne-Guigues basis}.

\subsection{Close-by-One \blue{and NextClosure}}
\label{sec:cbo}

We assume a closure operator $c$ on set $Y=\{1,2,\dots,n\}$.
Whenever we write about lower attributes or higher attributes,
we refer to the natural ordering of the numbers in $Y$.

We start the description of CbO with a basic algorithm for generating all closed sets
(Algorithm~\ref{alg:subsets}).
The basic algorithm
traverses the space of all subsets of $Y$, 
each subset is checked for closedness and is outputted.
This approach is quite inefficient as the number of closed subsets is typically significantly smaller than the number of all subsets.

\begin{algorithm}
\DontPrintSemicolon\LinesNumbered
\SetKwInOut{Input}{input}\SetKwInOut{Output}{output}
\SetKwProg{Fn}{def}{\string:}{}
\SetKwFunction{FRecurs}{GenerateFrom}%
\Fn{\FRecurs{$B$, $y$}}{
\Input{$B$ -- set of attributes\\ $y$ -- last added attribute}
\medskip
\If{$B = c(B)$}{%
  {\bf print}({$B$})
}
\medskip
\For{$i \in \{y+1,\dots,n\}$}{
$D \gets B \cup \{i\}$\;
\FRecurs{$D$, $i$}\;
}
{\bf return}
}
\medskip
\FRecurs{$\emptyset$, $0$}
\caption{Basic algorithm to enumerate closed subsets\label{alg:subsets}}
\end{algorithm}

The algorithm is given by a recursive procedure \texttt{GenerateFrom}, which accepts two arguments: 
\begin{itemize}
\item[$\bullet$] $B$ -- the set of attributes, from which new sets will be generated.
\item[$\bullet$] $y$ -- the auxiliary argument to remember the highest attribute in $B$.
\end{itemize}
The procedure first checks the input set $B$ for closedness and prints it if it is closed (lines 1,2). Then, for each
attribute $i$ higher than $y$:
\begin{itemize}
\item[$\bullet$]  a new set is generated by adding the attribute $i$ into the set $B$ (line~4);
\item[$\bullet$]  the procedure recursively calls itself to process the new set (line~5).
\end{itemize}
The procedure is initially called with an empty set and zero as its arguments.

\medskip

\begin{figure}
\begin{center}

\begin{center}
\begin{tikzpicture}[xscale=1.5]
\node (0L) at (1,0) {$\emptyset$};
\node[circle,draw] (0) at (1,0) {~~~};

\node (2L) at (-2,-1) {1};
\node[circle,draw] (2) at (2L) {~~~};

\node (3L) at (0,-1) {2};
\node[circle,draw] (3) at (3L) {~~~};

\node (4L) at (2,-1) {3};
\node[circle,draw] (4) at (4L) {~~~};

\node (5L) at (4,-1) {4};
\node[circle,draw] (5) at (5L) {~~~};

\node (45L) at (2,-2) {4};
\node[circle,draw] (45) at (45L) {~~~};

\node (34L) at (-.5,-2) {3};
\node[circle,draw] (34) at (34L) {~~~};

\node (35L) at (.5,-2) {4};
\node[circle,draw] (35) at (35L) {~~~};

\node (345L) at (-.5,-3) {4};
\node[circle,draw] (345) at (345L) {~~~};

\begin{scope}[xshift=-3cm,yshift=-1cm]

\node (23L) at (0,-1) {2};
\node[circle,draw] (23) at (23L) {~~~};

\node (24L) at (1,-1) {3};
\node[circle,draw] (24) at (24L) {~~~};

\node (25L) at (2,-1) {4};
\node[circle,draw] (25) at (25L) {~~~};

\node (245L) at (1,-2) {4};
\node[circle,draw] (245) at (245L) {~~~};

\node (234L) at (-.5,-2) {3};
\node[circle,draw] (234) at (234L) {~~~};

\node (235L) at (.5,-2) {4};
\node[circle,draw] (235) at (235L) {~~~};

\node (2345L) at (-.5,-3) {4};
\node[circle,draw] (2345) at (2345L) {~~~};
\end{scope}

\draw (0) -- (2);
\draw (0) -- (3);
\draw (0) -- (4);
\draw (0) -- (5);

\draw (2) -- (23);
\draw (2) -- (24);
\draw (2) -- (25);

\draw (3) -- (34);
\draw (3) -- (35);

\draw (4) -- (45);

\draw (23) -- (234);
\draw (23) -- (235);

\draw (24) -- (245);
\draw (34) -- (345);

\draw (234) -- (2345);

\node[fill=white,inner sep=1.5pt] (0L) at (0.north west) {\tiny 1};
\node[fill=white,inner sep=1.5pt] (2L) at (2.north west) {\tiny 2};
\node[fill=white,inner sep=1.5pt] (3L) at (3.north west) {\tiny 10};
\node[fill=white,inner sep=1.5pt] (4L) at (4.north west) {\tiny 14};
\node[fill=white,inner sep=1.5pt] (5L) at (5.north west) {\tiny 16};
\node[fill=white,inner sep=1.5pt] (45L) at (45.north west) {\tiny 15};
\node[fill=white,inner sep=1.5pt] (34L) at (34.north west) {\tiny 11};
\node[fill=white,inner sep=1.5pt] (35L) at (35.north west) {\tiny 13};
\node[fill=white,inner sep=1.5pt] (345L) at (345.north west) {\tiny 12};
\node[fill=white,inner sep=1.5pt] (23L) at (23.north west) {\tiny 3};
\node[fill=white,inner sep=1.5pt] (24L) at (24.north west) {\tiny 7};
\node[fill=white,inner sep=1.5pt] (25L) at (25.north west) {\tiny 9};
\node[fill=white,inner sep=1.5pt] (245L) at (245.north west) {\tiny 8};
\node[fill=white,inner sep=1.5pt] (234L) at (234.north west) {\tiny 4};
\node[fill=white,inner sep=1.5pt] (235L) at (235.north west) {\tiny 6};
\node[fill=white,inner sep=1.5pt] (2345L) at (2345.north west) {\tiny 5};

\draw[dotted,-latex] (0L) -- (2L);
\draw[dotted,-latex] (2L) -- (23L);
\draw[dotted,-latex] (23L) -- (234L);
\draw[dotted,-latex] (234L) -- (2345L);
\draw[dotted,-latex] (2345L) -- (235L);
\draw[dotted,-latex] (235L) -- (24L);
\draw[dotted,-latex] (24L) -- (245L);
\draw[dotted,-latex] (245L) -- (25L);
\draw[dotted,-latex] (25L) -- (3L);
\draw[dotted,-latex] (3L) -- (34L);
\draw[dotted,-latex] (34L) -- (345L);
\draw[dotted,-latex] (345L) -- (35L);
\draw[dotted,-latex] (35L) -- (4L);
\draw[dotted,-latex] (4L) -- (45L);
\draw[dotted,-latex] (45L) -- (5L);



\end{tikzpicture}
\end{center}

%
%
%
%
%
%
%
%
%
%
%
%
%
%
%
%
%
%
%
%
%
%
  
\end{center}
\caption{Tree of all subsets of $\{1,2,3,4\}$. Each node represents a unique set containing all elements in the path from the node to the root. 
The dotted arrows and small numbers represent the sweep performed by the CbO algorithm.
\label{fig:tree}}
\end{figure}
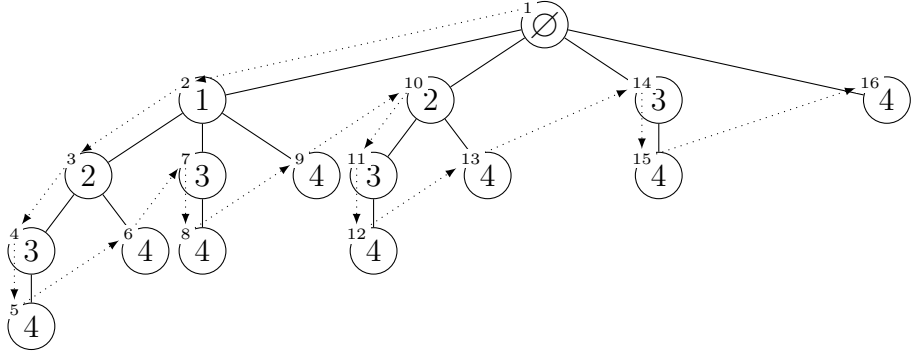

The basic algorithm represents a depth-first sweep through the tree of all subsets of $Y$ (see Fig.\,\ref{fig:tree}) and printing the closed ones.

In the tree of all subsets (Fig.\,\ref{fig:tree}), 
each node is a superset of its predecessors.
We can use the closure operator ${}^{\down\up}$ to skip non-closed sets. In other words, to make jumps in the tree to closed sets only.
\blue{CbO can be seen as the basic algorithm with closure jumps:}
instead of simply adding an element to generate a new subset
\[
D \gets
B \cup \{i\},
\]
CbO adds the element and then closes the set
\begin{equation}
\label{eq:D}
D \gets
c(B \cup \{i\}).
\end{equation}
We need to distinguish the two outcomes of the closure \eqref{eq:D}.
Either 
\begin{itemize}
\item
the closure contains some attributes lower than $i$ which are not included in $B$, i.e.
\[ 
D_i \neq B_i
\]
where
$D_i = D \cap \{1, \dots, i-1\},\,B_i = B \cap \{1, \dots, i-1\}$;
\item
or it does not, and we have
\[ 
D_i = B_i.
 \]
\end{itemize}
The jumps with $D_i \neq B_i$ are not desirable because they land on
a closed set which was already processed or will be processed later (depending on the direction of the sweep). 
CbO does not perform 
such jumps.
The check of the condition $D_i = B_i$ is called a {\em canonicity test}.

\RestyleAlgo{ruled}
\begin{algorithm}
\DontPrintSemicolon\LinesNumbered
\SetKwInOut{Input}{input}\SetKwInOut{Output}{output}
\SetKwProg{Fn}{def}{\string:}{}
\SetKwFunction{FRecurs}{CbOStep}%

\Fn{\FRecurs{$B$, $y$}}{
\Input{$B$ -- closed set\\ $y$ -- last added attribute}

\vspace{.25cm}

{\bf print}({$B$})\;

\medskip
\For{$i \in \{y+1,\dots, n \} \,\setminus\, B$}{
$D \gets c(B \cup \{i\})$\;
\If{$D_i = B_i$}{
\FRecurs{$D$, $i$}\;
}
}
}
\medskip

\FRecurs{$c(\emptyset)$, $0$}

\caption{Close-by-One\label{alg:cbo}}
\end{algorithm}

One can see the pseudocode of CbO in Algorithm~\ref{alg:cbo}.

We describe the differences from the basic algorithm:
\begin{itemize}
\item
The argument $B$ is a closed set, therefore, the procedure \texttt{GenerateFrom}
can print it directly without testing (line 1). 
\item In the loop, we skip elements already present in $B$ (line 2).
\item The recursive invocation is made only if the new closed set $D$ passes the canonicity test (lines 3,4).
\item
\green{The initial invocation is made with the smallest closed set $c(\emptyset)$ instead of the empty set.}
\end{itemize}

\blue{
The algorithm NextClosure \cite{GaWi:FCA} is another algorithm for enumerating closed sets.

NextClosure is represented by the procedure
\texttt{NextClosure} (Algorithm~\ref{alg:nc})
which accepts a closed set $B'$
and returns another closed set, which is the lectic successor of the input set.

It starts with a set $B$ containing all attributes from $B'$.
It processes attributes in $Y$ in descending order (line 2). 
\begin{itemize}
\item[1.]
If the processed attribute
is in $B$, it removes it (lines 3,4);
\item[2.]
If the processed attribute is not in $B$, it computes the closure $D$ of $B\cup \{i\}$ (lines 5,6);
\end{itemize}
Note that the above effectively increases the binary number corresponding to the characteristic vector
of $B$ by one and closes it; this corresponds to the description of the lectic order via binary numbers.
Then, the set $D$ is tested for canonicity the same way as in CbO.
If $D$ passes the test, it is returned as the result (line 7).
Otherwise, we continue processing the other attributes.
If we exhaust all attributes, we return $Y$ as the lectically last closed set.
\medskip

To enumerate all formal concepts, the NextClosure algorithm starts with the least closed set $c(\emptyset)$ and in consecutive steps applies this procedure to obtain the next formal concepts. The algorithm stops if $Y$ is obtained.

}

\begin{algorithm}
\DontPrintSemicolon\LinesNumbered
\SetKwInOut{Input}{input}\SetKwInOut{Output}{output}
\SetKwProg{Fn}{def}{\string:}{}
\SetKwFunction{NextClosure}{NextClosure}
\blue{
\Fn{\NextClosure{$B'$}}{
\Input{
    $B'$ -- set of attributes}
$B \gets B'$\;
\For{{\bf all} $i \in Y$ (in descending order)}{%
\eIf{$i \in B$}{$B\gets B \setminus \{i\}$}{
$D \gets c(B \cup \{i\})$\;
\lIf{$B_i = D_i$}{
{\bf return} $D$
}}}
{\bf return} $Y$
}
}
\caption{NextClosure\label{alg:nc}}
\end{algorithm}

\medskip
\blue{
NextClosure can be seen as an iterative 
version of CbO with the right depth-first sweep through the 
tree of all subsets. From this point of view, 
the above item 1.  
is equivalent to backtracking in the tree of all subsets, 
and item 2. is CbO's adding and closing. 
The consequent test of canonicity is the same as in CbO.
}
\subsection{LinClosure}
\label{sec:linclosure}

LinClosure (Algorithm~\ref{alg:linclosure}) \cite{beeri1979computational,maier1983theory} accepts a set $B$ of attributes for which it computes
the ${\cal T}$-closure 
$c_{\cal T}(B)$.
The theory $\cal T$ is considered to be a global variable.
It starts with a set $D$ containing all elements of $B$ (line 1).
If there is an attribute implication in $\cal T$ with an empty left side, the $D$ is united with its right side (lines 2,3).
LinClosure associates a counter $count[L \Impl R]$ with each $L \Impl R \in {\cal T}$
initializing it with the size $|L|$
of its left side (lines 4,5).  
Also, each attribute $y \in Y$ is linked to a list of the attribute implications that have $y$ 
in their left sides (lines 6,7).\footnote{This needs to be done just once and it is usually done outside the LinClosure procedure.}
Then, the set $Z$ of attributes to be processed is initialized as a copy of the set $D$ (line 8).
While there are attributes in $Z$,
the algorithm chooses one of them ($\min$ in the pseudocode, line 10), removes it from $Z$
(line 11) and decrements counters of all attribute implication linked to it (lines 12,13).
If the counter of any attribute implication $L \Impl R$ is decreased to 0, new 
attributes from $R$ are added to $D$ and to $Z$.

\begin{algorithm}
\DontPrintSemicolon\LinesNumbered
\SetKwInOut{Input}{input}\SetKwInOut{Output}{output}
\SetKwProg{Fn}{def}{\string:}{}
\SetKwFunction{LinClosure}{LinClosure}
\SetKwFunction{Min}{Min}

\Fn{\LinClosure{$B$
}}{
  \Input{
    $B$ -- set of attributes}
\medskip

$D \gets B$

\If{$\exists \emptyset \Impl R \in {\cal T} \text{ for some $R$}$}{
$D \gets D \cup R$
}

\For{ {\bf all} $L \Impl R \in {\cal T}$}
{
$count[L \Impl R] \gets |L|$\;
\For{ {\bf all} $a \in L$}{
{\bf add} $L \Impl R$ {\bf to} $list[a]$\;
}
}
\medskip
$Z \gets D$\;
\medskip

\While{$Z \neq \emptyset$}{
$m \gets \min(Z)$\;
$Z \gets Z \setminus \{ m \}$\;
\For{ {\bf all} $L \Impl R \in list[m]$}{
$count[L\Impl R] \gets count[L \Impl R] -1$\;
\If{$count[L \Impl R] = 0$}{
$add \gets R \setminus D$\;
$D \gets D \cup add$\;
$Z \gets Z \cup add$\;
}}
}
{\bf return} $D$\;
}
\caption{LinClosure \label{alg:linclosure}}
\end{algorithm}




We are going to use the algorithm LinClosure in CbO.  
CbO drops the resulting closed set if it fails
the canonicity test (Algorithm~\ref{alg:cbo}, lines~4,5).
Therefore, we can introduce a feature \blue{-- early stop --} which
stops the computation whenever an attribute which would
cause the fail is added into the set. 
To do that, we add a new input argument, $y$, having the same role as in CbO; i.e. 
the last attribute added into the set (Algorithm~\ref{alg:linclosure2}). 
Then, whenever new attributes are added to the set, we check whether 
any of them is lower than $y$. If so, we stop the procedure and return 
 information that the canonicity test would fail (lines 16--17).\footnote{This feature is also utilized in \cite{bazhanov2014}.}

\blue{
In the pseudocode of LinClosure with an early stop
(Algorithm~\ref{alg:linclosure2}), we also removed the two lines which handled
the case for the attribute implication in $\cal T$ with an empty left side
(Algorithm~\ref{alg:linclosure}, lines~2,3).
In Section~\ref{sec:nc_improv}, we introduce an improvement for CbO which
makes the two lines superfluous.
}

\begin{algorithm}
\DontPrintSemicolon\LinesNumbered
\SetKwInOut{Input}{input}\SetKwInOut{Output}{output}
\SetKwProg{Fn}{def}{\string:}{}
\SetKwFunction{LinClosure}{LinClosureES}
\SetKwFunction{Min}{Min}

\Fn{\LinClosure{$B$, $y$
}}{
  \Input{
    $B$ -- set of attributes\\
$y$ -- last attribute added to $B$}
\medskip

$D \gets B$\;

\smallskip

\If{$\exists \emptyset \Impl R \in {\cal T} \text{ for some $R$}$}{
$D \gets D \cup R$
}

\For{ {\bf all} $L \Impl R \in {\cal T}$}
{
$count[L \Impl R] \gets |L|$\;
\For{ {\bf all} $a \in L$}{
{\bf add} $L \Impl R$ {\bf to} $list[a]$\;
}
}
\medskip
$Z \gets D$\;
\medskip

\While{$Z \neq \emptyset$}{
$m \gets \min(Z)$\;
$Z \gets Z \setminus \{ m \}$\;
\For{ {\bf all} $L \Impl R \in list[m]$}{
$count[L\Impl R] \gets count[L\Impl R] -1$\;
\If{$count[L \Impl R] = 0$}{
$add \gets R \setminus D$\;
\eIf{$\min(add) < y$}{
{\bf return} fail\;
}{
$D \gets D \cup add$\;
$Z \gets Z \cup add$\;
}}}
}
{\bf return} $D$\;
}
\caption{LinClosure with an early stop\label{alg:linclosure2}}
\end{algorithm}

\blue{
\subsection{Wild's closure}
\label{sec:wildclosure}

For the sake of completeness, we also describe Wild's closure \cite{wild1995computations}.
Our algorithm does not use this closure; however, algorithms $\mathrm{NC}3$ and $\mathrm{NC}^+3$, which we use in the experimental evaluation (Section~\ref{sec:ee}),
do so.

Wild's closure (Algorithm~\ref{alg:wildclosure}) 
accepts a set $B$ of attributes for which it computes
the ${\cal T}$-closure 
$c_{\cal T}(B)$.
The theory $\cal T$ is considered to be a global variable.

It starts with a set $D$ containing all elements of $B$ (line 1).
First, it handles the case for attribute implication with an empty left side, the same way that LinClosure does 
(lines 2,3).
Wild's closure maintains implication lists, similarly to LinClosure (lines 4-6).
It keeps a set $\cal N$ of current attribute implications, initially equal to $\cal T$ (line 7).
It uses the attribute lists to find a subset 
${\cal N}_1 \subseteq \cal N$ of implications whose left-hand
side has an attribute not occurring in $D$ (line 10).
It uses the rest ${\cal N}\,\setminus\,{\cal N}_1$ of implications
to extend $D$. If $D$ is extended, the process is repeated for ${\cal N}_1$ 
being the set of current implications (loop at lines 8-15). Otherwise $D$ is the
resulting set and is returned (line 16).



\begin{algorithm}
\blue{
\DontPrintSemicolon\LinesNumbered
\SetKwInOut{Input}{input}\SetKwInOut{Output}{output}
\SetKwProg{Fn}{def}{\string:}{}
\SetKwFunction{WildClosure}{WildClosure}

\Fn{\WildClosure{$B$}}{
\Input{
    $B$ -- set of attributes}
\SetKwRepeat{RepeatUntil}{repeat}{until}
$D \gets B$\;

\If{$\exists \emptyset \Impl R \in {\cal T} \text{ for some $R$}$}{
$D \gets D \cup R$
}

\For{{\bf all} $L \Impl R \in {\cal T}$}{
    \For{$a \in L$}{{\bf add} $L \Impl R$ {\bf to} $list[a]$}
}
${\cal N} \gets {\cal T}$

\RepeatUntil{$stable$}{
$stable \gets true$\;
${\cal N}_1 \gets \bigcup_{a \notin D} list[a]$\;
\For{{\bf all} $L \Impl R \in {\cal N} \,\setminus\, {\cal N}_1$}{
$D \gets D \cup R$\;
$stable \gets false$;
}
${\cal N} \gets {\cal N}_1$\;
}
{\bf return} $D$\;
}
}
\caption{Wild's closure\label{alg:wildclosure}}
\end{algorithm}

}

\section{LinCbO: CbO-based algorithm for computation of the Duquenne-Guigues basis}
\label{sec:alg}
In this section, we describe the algorithm LinCbO. Its foundation is CbO (Algorithm~\ref{alg:cbo})
with LinClosure (Algorithm~\ref{alg:linclosure}).
We explain changes in the CbO algorithm: a change of sweep order makes the algorithms work, and the rest of the changes improve efficiency of the algorithms.

\subsection{Sweep order}
\label{sec:sweep_order}
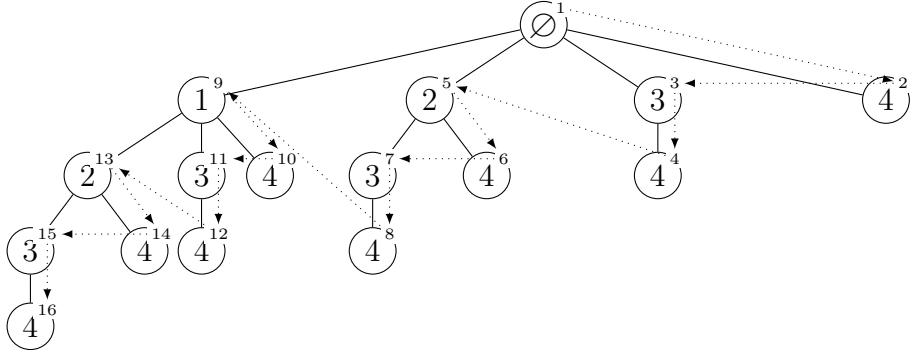
\begin{figure}
\begin{center}


\begin{center}
\begin{tikzpicture}[xscale=1.5]
\node (0L) at (1,0) {$\emptyset$};
\node[circle,draw] (0) at (1,0) {~~~};

\node (2L) at (-2,-1) {1};
\node[circle,draw] (2) at (2L) {~~~};

\node (3L) at (0,-1) {2};
\node[circle,draw] (3) at (3L) {~~~};

\node (4L) at (2,-1) {3};
\node[circle,draw] (4) at (4L) {~~~};

\node (5L) at (4,-1) {4};
\node[circle,draw] (5) at (5L) {~~~};

\node (45L) at (2,-2) {4};
\node[circle,draw] (45) at (45L) {~~~};

\node (34L) at (-.5,-2) {3};
\node[circle,draw] (34) at (34L) {~~~};

\node (35L) at (.5,-2) {4};
\node[circle,draw] (35) at (35L) {~~~};

\node (345L) at (-.5,-3) {4};
\node[circle,draw] (345) at (345L) {~~~};

\begin{scope}[xshift=-3cm,yshift=-1cm]

\node (23L) at (0,-1) {2};
\node[circle,draw] (23) at (23L) {~~~};

\node (24L) at (1,-1) {3};
\node[circle,draw] (24) at (24L) {~~~};

\node (25L) at (1.6,-1) {4};
\node[circle,draw] (25) at (25L) {~~~};

\node (245L) at (1,-2) {4};
\node[circle,draw] (245) at (245L) {~~~};

\node (234L) at (-.5,-2) {3};
\node[circle,draw] (234) at (234L) {~~~};

\node (235L) at (.5,-2) {4};
\node[circle,draw] (235) at (235L) {~~~};

\node (2345L) at (-.5,-3) {4};
\node[circle,draw] (2345) at (2345L) {~~~};
\end{scope}

\draw (0) -- (2);
\draw (0) -- (3);
\draw (0) -- (4);
\draw (0) -- (5);

\draw (2) -- (23);
\draw (2) -- (24);
\draw (2) -- (25);

\draw (3) -- (34);
\draw (3) -- (35);

\draw (4) -- (45);

\draw (23) -- (234);
\draw (23) -- (235);

\draw (24) -- (245);
\draw (34) -- (345);

\draw (234) -- (2345);

\node[fill=white,inner sep=1.5pt] (0L) at (0.north east) {\tiny 1};
\node[fill=white,inner sep=1.5pt] (2L) at (2.north east) {\tiny 9};
\node[fill=white,inner sep=1.5pt] (3L) at (3.north east) {\tiny 5};
\node[fill=white,inner sep=1.5pt] (4L) at (4.north east) {\tiny 3};
\node[fill=white,inner sep=1.5pt] (5L) at (5.north east) {\tiny 2};
\node[fill=white,inner sep=1.5pt] (45L) at (45.north east) {\tiny 4};
\node[fill=white,inner sep=1.5pt] (34L) at (34.north east) {\tiny 7};
\node[fill=white,inner sep=1.5pt] (35L) at (35.north east) {\tiny 6};
\node[fill=white,inner sep=1.5pt] (345L) at (345.north east) {\tiny 8};
\node[fill=white,inner sep=1.5pt] (23L) at (23.north east) {\tiny 13};
\node[fill=white,inner sep=1.5pt] (24L) at (24.north east) {\tiny 11};
\node[fill=white,inner sep=1.5pt] (25L) at (25.north east) {\tiny 10};
\node[fill=white,inner sep=1.5pt] (245L) at (245.north east) {\tiny 12};
\node[fill=white,inner sep=1.5pt] (234L) at (234.north east) {\tiny 15};
\node[fill=white,inner sep=1.5pt] (235L) at (235.north east) {\tiny 14};
\node[fill=white,inner sep=1.5pt] (2345L) at (2345.north east) {\tiny 16};

\draw[dotted,-latex] (0L) -- (5L);
\draw[dotted,-latex] (5L) -- (4L);
\draw[dotted,-latex] (4L) -- (45L);
\draw[dotted,-latex] (45L) -- (3L);
\draw[dotted,-latex] (3L) -- (35L);
\draw[dotted,-latex] (35L) -- (34L);
\draw[dotted,-latex] (34L) -- (345L);
\draw[dotted,-latex] (345L) -- (2L);
\draw[dotted,-latex] (2L) -- (25L);
\draw[dotted,-latex] (25L) -- (24L);
\draw[dotted,-latex] (24L) -- (245L);
\draw[dotted,-latex] (245L) -- (23L);
\draw[dotted,-latex] (23L) -- (235L);
\draw[dotted,-latex] (235L) -- (234L);
\draw[dotted,-latex] (234L) -- (2345L);



\end{tikzpicture}
\end{center}

%
%
%
%
%
%
%
%
%
%
%
%
%
%
%
%
%
%
%
%
%
%
  
\end{center}
\caption{Tree of all subsets of $\{1,2,3,4\}$. Each node represents a unique set containing all elements in the path from the node to the root. 
The dotted arrows and small numbers represent the sweep performed by the CbO algorithm with right depth-first sweep.
\label{fig:tree2}}
\end{figure}

In the previous section, we presented CbO as the left first sweep through the tree
of all subsets. This is how it is usually described. 
In ordinary settings, there is no need to follow a particular order of sweep. 
However, our aim is to compute intents and pseudo-intents using the closure
operator $\tilde{c}_{\cal T}$~\eqref{eq:closure_operator}\blue{, or more exactly, closure operator 
${c}_{\cal S}$ for 
${\cal S} \subseteq {\cal T}$ 
satisfying \eqref{eq:huhuhu}}. For this, we need to utilize
an order which extends the subsethood, i.e. \eqref{eq:subsethood}. The right depth-first
sweep through the tree of all subsets satisfies this condition (see Fig.\,\ref{fig:tree2}).
Observe that with the right depth-first sweep, we obtain exactly the lectic order,
i.e. the same order in which NextClosure explores the search space.

\subsection{NextClosure's improvements}
\label{sec:nc_improv}

The following improvements were introduced to NextClosure \cite{bazhanov2014} and the incremental approach \cite{obiedkov2007attribute} for computation of pseudo-intents. We incorporated them to the CbO algorithm.

After the algorithm computes $B^\bullet$, the implication $B^\bullet \to B^{\down\up}$
is added to $\cal T$, provided $B^\bullet$ is a pseudo-intent, i.e. $B^\bullet \neq B^{\down\up}$.

Note that there exists the smallest $\tilde{c}_{\cal T}$-closed set larger than $B^\bullet$ and it is
the intent $B^{\bullet\down\up}$ ($= B^{\down\up}$). Consider the following two cases:

\begin{itemize}
\item[(o1)] This intent satisfies the canonicity test, i.e. $(B^{\down\up})_y=(B^{\bullet})_y$, where $y$ is the last added attribute to $B$. Then we can jump to this intent.

\item[(o2)] This intent does not satisfy the canonicity test. Thus, we can leave the present subtree.
\end{itemize}

%

Now, let us describe the first version of LinCbO  (Algorithm~\ref{alg:dgcbo1}), which includes the above discussed improvements.\footnote{\green{As CbO with right depth-first sweep can be considered a recursive  NextClosure, this version of LinCbO can be considered a recursive version of the corresponding algorithm from \cite{bazhanov2014} (denoted NC$^+$2 later in this paper).}}

The procedure \texttt{LinCbO1Step} works with the following global variables: an initially empty theory $\cal T$ and 
an initially empty list of attribute implications for each attribute.
\texttt{LinCbO1Step} accepts two arguments: a set $B$ of attributes and the last attribute 
$y$ added to $B$. The set $B$ is not generally closed (which was the case in Algorithm~\ref{alg:cbo}).

The procedure first applies LinClosure with an early stop (Algorithm~\ref{alg:linclosure2}) to compute $B^\bullet$ (line~1).
If $B^\bullet$ fails the canonicity test (recall that the canonicity test is incorporated in LinClosure with an early stop), the procedure stops (lines~2,3).
Then, the procedure computes $B^{\bullet\down\up}$ to check whether $B^\bullet$ is an intent or pseudo-intent (line~4).
If it is a pseudo-intent, a new attribute implication $B^\bullet \Impl B^{\bullet\down\up}$
is added to the initially empty theory $\cal T$ (line 5). 
For each attribute in $B^\bullet$, we update its list by adding the new attribute implication (lines 6 and 7).

Now, as we computed the intent $B^{\bullet\down\up}$, we can apply (o1) or (o2) based on the result of  the canonicity test $(B^{\bullet\down\up})_y = (B^\bullet)_y$ (line 8) -- either we call \texttt{LinCbO1Step} for $B^{\bullet\down\up}$ (line 9) or end the procedure.
If $B^\bullet$ is an intent, we recursively call \texttt{LinCbO1Step} for all sets $B^\bullet \cup \{ i \}$ 
where $i$ is higher than the last added attribute $y$ and is not already present in $B^\bullet$.
To have lectic order, we make the recursive calls in the descending order of $i$s.

The procedure \texttt{LinCbO1Step} is initially called with empty set of attributes and zero representing an invalid last added attribute.

\begin{algorithm}
\caption{LinCbO1 (CbO for the Duquenne-Guigues basis, first version) \label{alg:dgcbo1}}
\DontPrintSemicolon\LinesNumbered
\SetKwFunction{CbO}{LinCbO1Step}%
\SetKwFunction{Closure}{LinClosureES}
\SetKwProg{Fn}{def}{\string:}{}
\SetKwInOut{Input}{input}\SetKwInOut{Output}{output}

\medskip

$\cal T \gets \emptyset$\;
$list[i] \gets \emptyset$ {\bf for each } $i\in Y$\;
\medskip

\Fn{\CbO{$B$, $y$}}{
  \Input{
    $B$ -- set of attributes\\
    $y$ -- last attribute added to $B$}

\medskip
$B^\bullet \gets \Closure(B,y)$\;
\If{$B^\bullet$ is fail} {
return\;
}
\eIf{$B^\bullet \neq B^{\bullet\down\up}$}{

  ${\cal T} \gets {\cal T} \cup \{ B^\bullet\Impl B^{\bullet\down\up} \}$\;
      
  \For{$i \in B^\bullet $}{
    $list[i] \gets list[i] \cup \{ B^\bullet\Impl B^{\bullet\down\up} \}$\;
  }

  \If{$
  (B^{\bullet\down\up})_y = (B^\bullet)_y
  $}{
  	\CbO{$B^{\bullet\down\up},y$}\;  
  }
}{
  \For{$i$ {\bf from} $n$ {\bf down to} $y+1$, $i \notin B^{\bullet}$}{
    \CbO{$B^{\bullet} \cup \{i\}, i$}
  }
}}
\medskip

\CbO{$\emptyset,0$}\; 
\end{algorithm}

\medskip

\blue{
Now we can explain why we removed the part of the code of LinClosure
which handles the case $\emptyset \Impl R \in \cal T$ (Algorithm~\ref{alg:linclosure}, lines~2,3) from
LinClosure with an early stop.
The presence of $\emptyset \Impl R$ in $\cal T$ means that $\emptyset$ is a pseudo-intent.
This pseudo-intent is generated by the initial invocation of \texttt{LinCbO1Step}.
Since for the initial invocation, we have $y=0$, the intent $B^{\bullet\down\up} = \emptyset^{\down\up} = R$
trivially satisfies the condition $R_y = \emptyset_y$ (Algorithm~\ref{alg:dgcbo1}, line 8)
and \texttt{LinCbO1Step} is invoked with this intent (Algorithm~\ref{alg:dgcbo1}, line 9). 
Consequently, all the processed sets are supersets of $R$, and therefore the union with $R$ 
(Algorithm~\ref{alg:linclosure}, line 3) does nothing.
}

%
%
%

\subsection{LinClosure with reused counters}
\label{sec:imlc}

Consider theory ${\cal T}'$ and theory ${\cal T}$ which emerges by adding new attribute implications
to ${\cal T'}$, i.e. ${\cal T}' \subseteq {\cal T}$.
When we compute ${\cal T}'$-closure ${B'}$, we can store values of the attribute
counters at the end of the LinClosure procedure.
Later, when we compute ${\cal T}$-closure of a superset $B$ of $B'$, we can initialize the attribute
counters of implications from ${\cal T}'$ to the stored values instead of the antecedent sizes.
Attribute counters for new implications, i.e. those in ${\cal T}' \setminus {\cal T}$, are initialized the usual way. Then, we handle only the new attributes, that is those in $B\setminus B'$.

We can improve the LinClosure accordingly (Algorithm~\ref{alg:linclosure3}). We describe only the differences from LinClosure with an early stop (Algorithm~\ref{alg:linclosure2}). It accepts two additional arguments: $Z'$ -- the set of new attributes, i.e, those which were not in the ${\cal T}$-closed subset from which we reuse the counters; and $prevCount$ -- the previous counters to be reused.
We copy the previous counters and new attributes $Z'$ to local variables (lines 2,3). Furthermore, we add new attribute implications (lines 4,5).

\begin{algorithm}
\DontPrintSemicolon\LinesNumbered
\SetKwInOut{Input}{input}\SetKwInOut{Output}{output}
\SetKwProg{Fn}{def}{\string:}{}
\SetKwFunction{LinClosure}{LinClosureRC}
\SetKwFunction{Min}{Min}

\Fn{\LinClosure{$B$, $y$, $Z'$, $prevCount$}}{
  \Input{
    $B$ -- set of attributes to be closed\\
    $y$ -- last attribute added to $B$\\
    $Z'$ -- set of new attributes\\
    $\mathit{prevCount}$ -- previous attribute counters from computation $B \setminus Z$}
\medskip

$D \gets$ $B$\;

$\mathit{count} \gets$ copy of $\mathit{prevCount}$\;
$Z \gets Z'$

\For{$L \Impl R \in {\cal T}$ not counted in $\mathit{prevCount}$}{
  $count[L \Impl R] \gets |L \setminus B|$\;
}

\smallskip

\While{$Z \neq \emptyset$}{
  $m \gets \min(Z)$
  
  $Z \gets Z \setminus \{m\}$

  \For{$L \Impl R \in list[m]$}{
    $\mathit{count}[L\Impl R]\gets \mathit{count}[L\Impl R]-1$
    
    \If{$\mathit{count}[L \Impl R]=0$}{
      $\mathit{add}\gets R \setminus D$

      \If{$\min(\mathit{add}) < y$}{
        {\bf return} fail
      }
      $D \gets D \cup \mathit{add}$
      
      $Z \gets Z \cup \mathit{add}$
    }
    
  }}
  {\bf return } $\tu{D, \mathit{count}}$

}
\caption{LinClosure with reused counters \label{alg:linclosure3}}
\end{algorithm}

Note, that in CbO we always make the recursive invocations for supersets of the current set (see
Algorithm~\ref{alg:dgcbo1}, lines 9 and 12).
Therefore, 
we can easily utilize the LinClosure with reused counters in LinCbO (Algorithm~\ref{alg:dgcbo2}).
The only difference from the first version (Algorithm~\ref{alg:dgcbo1}) is that the procedure $\texttt{LinCbOStep}$
accepts two additional arguments, which are passed to procedure \texttt{LinClosureRC} 
(line 1). The two arguments
are: the set of new attributes and the previous attribute counters (both initially empty).
Recall that the attribute counters are modified by LinClosure.
The corresponding arguments are also passed to the recursive invocations of \texttt{LinCbOStep}
(lines 9 and 12).

\begin{algorithm}
\caption{LinCbO  (CbO for the Duquenne-Guigues basis, final version\label{alg:dgcbo2})}
\DontPrintSemicolon\LinesNumbered
\SetKwFunction{CbO}{LinCbOStep}%
\SetKwFunction{Closure}{LinClosureRC}
\SetKwProg{Fn}{def}{\string:}{}
\SetKwInOut{Input}{input}\SetKwInOut{Output}{output}

\medskip

$\cal T \gets \emptyset$\;
$list[i] \gets \blue{\emptyset}$ {\bf for each } $y\in Y$\;

\medskip

\Fn{\CbO{$B$, $y$, $Z$, $prevCount$}}{
  \Input{
    $B$ -- set of attributes\\
    $y$ -- last attribute added to $B$\\
    $Z$ -- set of new attributes\\
    $prevCount$ -- attribute counters}

\medskip
$\tu{B^\bullet, count} \gets \Closure(B,y,Z,prevCount)$\;
\If{$B^\bullet$ is fail} {
{\bf return}\;
}
\eIf{$B^\bullet \neq B^{\bullet\down\up}$}{

  ${\cal T} \gets {\cal T} \cup \{ B^\bullet\Impl B^{\bullet\down\up} \}$\;
      
  \For{$i \in B^\bullet $}{
    $list[i] \gets list[i] \cup \{ B^\bullet\Impl B^{\bullet\down\up} \}$\;
  }

  \If{$
  (B^{\bullet\down\up})_y = (B^\bullet)_y
  $}{
  	\CbO{$B^{\bullet\down\up},y, B^{\bullet\down\up}\setminus B^\bullet$, count}\;  
  }
}{
  \For{$i$ {\bf from} $n$ {\bf down to} $y+1$, $i \notin B^{\bullet}$}{
    \CbO{$B^{\bullet} \cup \{i\}, i, \{i\}, count$}
  }
}}
\medskip

\CbO{$\emptyset,0,\emptyset,\emptyset$}\; 
\end{algorithm}

\section{Experimental Comparison}
\label{sec:ee}

We compare LinCbO with other algorithms, namely:
\begin{itemize}
\item
NextClosure with na\"ive closure (NC1), LinClosure (NC2), and Wild's closure (NC3).
\item
NextClosure$^+$, which is NextClosure with the improvements described in Section~\ref{sec:nc_improv},  with the same closures (NC$^+$1, NC$^+$2, NC$^+$3)\footnote{NextClosure and NextClosure$^+$ are called Ganter and Ganter$^+$ in \cite{bazhanov2014}.};
\item
attribute incremental approach \cite{obiedkov2007attribute}.
\end{itemize}
To achieve maximal fairness, we implemented LinCbO 
into the framework made
by Bazhanov \& Obiedkov \cite{bazhanov2014}\footnote{Available at \url{https://github.com/yazevnul/fcai}}. 
It contains implementations of all the listed algorithms.
In Section \ref{sec:ee1}, we also use the same datasets as used by Bazhanov and Obiedkov \cite{bazhanov2014}.

All experiments have been performed on a computer with 64\,GB RAM, two Intel Xeon CPU E5-2680 v2 (at 2.80\,GHz), Debian Linux 10, and GNU GCC 8.3.0.  
All measurements have been taken ten times and the mean value is presented.


\subsection{Batch 1: datasets used in \cite{bazhanov2014}}
\label{sec:ee1}

Bazhanov and Obiedkov \cite{bazhanov2014} use artificial datasets and datasets from UC Irvine Machine Learning Repository \cite{uci}.

The artificial datasets are named as $|X|$\texttt{x}$|Y|$\texttt{-}$d$, where $d$ is the number of 
attributes of each object; i.e. $|\{x\}^{\up}|=d$ for each $x\in X$. The attributes are assigned
to objects randomly, with exception \texttt{18x18-17}, where each object misses a different attribute 
(more exactly, the incidence relation is the inequality).

The datasets from UC Irvine Machine Learning Repository are:
\texttt{Breast-cancer},
\texttt{Breast-w},
\texttt{dbdata0},
\texttt{flare},
\texttt{Post-operative},
\texttt{spect},
\texttt{vote}, and
\texttt{zoo}. See Table~\ref{tab:small0} for properties of all the datasets.


\begin{table}
\setlength{\tabcolsep}{12pt}
\caption{Properties of the datasets in batch 1\label{tab:small0}}
\begin{tabular}{l||
r|r|r|r|r}
  dataset & {$|X|$} & {$|Y|$} & {$|I|$} & \# {intents} & \# {ps.intents} \\
\hline
\hline
\texttt{100x30-4}    &  100 &  30 &   400 &     307 &    557 \\
\texttt{100x50-4}    &  100 &  50 &   400 &     251 &   1115 \\
\texttt{10x100-25}  &   10 & 100 &   250 &     129 &    380 \\
\texttt{10x100-50}  &   10 & 100 &   500 &     559 &    546 \\
\texttt{18x18-17}   &   18 &  18 &   306 & 262,144 &      0 \\
\texttt{20x100-25}  &   20 & 100 &   500 &     716 &   2269 \\
\texttt{20x100-50}  &   20 & 100 &  1000 &  12,394 &   8136 \\
\texttt{50x100-10}  &   50 & 100 &   500 &     420 &   3893 \\
\texttt{900x100-4}  &  900 & 100 &  3600 &    2472 &   7994 \\\hline
\texttt{Breast-cancer}  &  286 &  43 &  2851 &    9918 &   3354 \\
\texttt{Breast-w}       &  699 &  91 &  6974 &    9824 & 10,666 \\
\texttt{dbdata0}        &  298 &  88 &  1833 &    2692 &   1920 \\
\texttt{flare}          & 1389 &  49 &18,062 &  28,742 &   3382 \\
\texttt{Post-operative} &   90 &  26 &   807 &    2378 &    619 \\
\texttt{spect}          &  267 &  23 &  2042 &  21,550 &   2169 \\
\texttt{vote}           &  435 &  18 &  3856 &  10,644 &    849 \\
\texttt{zoo}            &  101 &  28 &   862 &     379 &    141 \\
\end{tabular}
\end{table}


In batch 1, LinCbO computes the basis faster than the rest of algorithms;
 however in most cases the runtimes are very small and differences between
 them are negligible (see Table~\ref{tab:small}).



\begin{sidewaystable}
\setlength{\tabcolsep}{12pt}
\renewcommand{\bfseries}{\fontseries{b}\selectfont} 
\robustify\bfseries             
\newrobustcmd{\BB}{\bfseries}    

\caption{Runtimes in seconds of algorithms generating Duquenne-Guigues basis in batch 1.  \label{tab:small}}
\begin{tabular}{l||
S[table-format=4.3,detect-weight,mode=text]|
S[table-format=4.3,detect-weight,mode=text]|
S[table-format=4.3,detect-weight,mode=text]|
S[table-format=4.3,detect-weight,mode=text]|
S[table-format=4.3,detect-weight,mode=text]|
S[table-format=4.3,detect-weight,mode=text]|
S[table-format=4.3,detect-weight,mode=text]|
S[table-format=4.3,detect-weight,mode=text]}
Dataset       & {AttInc} & {NC1} & {NC2} & {NC3} & {NC$^+$1} & {NC$^+$2} & {NC$^+$3} & {LinCbO} \\
\hline
\hline
\texttt{100x30-4}      & 0.008 & 0.007 & 0.007 & 0.01  & 0.004 & 0.003 & 0.005 & \BB 0.002 \\
\texttt{100x50-4}      & 0.028 & 0.037 & 0.024 & 0.05  & 0.013 & 0.008 & 0.016 & \BB 0.005 \\
\texttt{10x100-25}     & 0.015 & 0.015 & 0.023 & 0.033 & 0.007 & 0.01  & 0.014 & \BB 0.004 \\
\texttt{10x100-50}     & 0.037 & 0.052 & 0.087 & 0.112 & 0.038 & 0.063 & 0.081 & \BB 0.015 \\
\texttt{18x18-17}      & 0.337 & 0.096 & 0.143 & 0.134 & 0.111 & 0.157 & 0.151 & \BB 0.148 \\
\texttt{20x100-25}     & 0.099 & 0.281 & 0.165 & 0.484 & 0.094 & 0.061 & 0.172 & \BB 0.026 \\
\texttt{20x100-50}     & 0.94  & 5.457 & 3.047 & 8.898 & 3.809 & 2.31  & 6.481 & \BB 0.675 \\
\texttt{50x100-5}      & 0.454 & 0.778 & 0.253 & 1.064 & 0.126 & 0.047 & 0.164 & \BB 0.029 \\
\texttt{900x100-4}     & 2.061 & 3.315 & 0.91  & 3.936 & 1.15  & 0.317 & 1.333 & \BB 0.172 \\\hline
\texttt{Breast-cancer} & 0.121 & 0.295 & 0.236 & 0.325 & 0.231 & 0.184 & 0.251 & \BB 0.055 \\
\texttt{Breast-w}      & 2.856 & 4.674 & 3.128 & 9.61  & 2.526 & 1.67  & 5.155 & \BB 0.516 \\
\texttt{dbdata0}       & 0.109 & 0.254 & 0.312 & 0.43  & 0.158 & 0.208 & 0.263 & \BB 0.049 \\
\texttt{flare}         & 0.622 & 1.006 & 1.865 & 1.813 & 0.92  & 1.661 & 1.624 & \BB 0.265 \\
\texttt{Post-operative}& 0.014 & 0.015 & 0.023 & 0.021 & 0.013 & 0.018 & 0.018 & \BB 0.009 \\
\texttt{spect}         & 0.142 & 0.407 & 0.584 & 0.397 & 0.388 & 0.556 & 0.377 & \BB 0.097 \\
\texttt{vote}          & 0.054 & 0.062 & 0.078 & 0.068 & 0.059 & 0.075 & 0.064 & \BB 0.024 \\
\texttt{zoo}           & 0.004 & 0.003 & 0.005 & 0.005 & \BB 0.002 & 0.004 & 0.004 & \BB 0.002 
\end{tabular}
\end{sidewaystable}

\subsection{Batch 2: our collection of datasets}
\label{sec:ee2}
As the runtimes in batch 1 often differ only in a few milliseconds, we tested the algorithm on larger datasets.
We used the following datasets from UC Irvine Machine Learning Repository \cite{uci}:
\begin{itemize}
\item
\texttt{crx} -- Credit Approval (37 rows containing a missing value were removed), 
\item
\texttt{shuttle} -- Shuttle Landing Control, 
\item
\texttt{magic} -- MAGIC Gamma Telescope, 
\item
\texttt{bikesharing\_(day|hour)} -- Bike Sharing Dataset, 
\item
\texttt{kegg} -- KEGG Metabolic Reaction Network -- Undirected. 
\end{itemize}

We 
binarized the datasets
using nominal (\texttt{nom}), ordinal (\texttt{ord}), and interordinal (\texttt{inter}) scaling,
where each numerical feature was scaled to $k$ attributes
with $k-1$ equidistant cutpoints. 
Categorical features were scaled nominally to a number 
of attributes corresponding to the number of categories.
After the binarization,
we removed full columns. Properties of the resulting datasets are shown in Table~\ref{tab:big0}.
The naming convention used in Table~\ref{tab:big0} (and Table~\ref{tab:big}) is the following:
$(\text{scaling})k(\text{dataset})$. For example, \texttt{inter10shuttle}
is the dataset `Shuttle Landing Control' interordinally scaled to 10, using 9 equidistant cutpoints.


\begin{sidewaystable}
\setlength{\tabcolsep}{12pt}
\caption{Properties of the datasets in batch 2\label{tab:big0}}
\begin{tabular}{l||
r|r|r|r|r}
  dataset & {$|X|$} & {$|Y|$} & {$|I|$} & \# {intents} & \# {ps.intents} \\
\hline
\hline
\texttt{inter10crx} & 653   & 139 &   40,170 & 10,199,818 & 20,108 \\
\texttt{inter10shuttle}        & 43,500 & 178 & 3,567,907 & 38,199,148 & 936 \\
\texttt{inter3magic}           & 19,020 & 52  &  399,432 &  1,006,553 & 4181 \\
\texttt{inter4magic}           & 19,020 & 72  &  589,638 & 24,826,749 & 21,058 \\
\texttt{inter5bike\_day}        & 731   & 93  &   24,650 &  3,023,326 & 20,425 \\
\texttt{inter5crx}             & 653   & 79  &   20,543 &   348,428 & 3427 \\
\texttt{inter5shuttle}         & 43,500 & 88  & 1,609,510 &   333,783 & 346 \\
\texttt{inter6shuttle}         & 43,500 & 106 & 2,002,790 &   381,636 & 566 \\\hline
\texttt{nom10bike\_day}        &   731 & 100 &    9293 &    52,697 & 29,773 \\
\texttt{nom10crx}              &   653 &  85 &    8774 &    51,078 & 6240 \\
\texttt{nom10magic}            & 19,020 & 102 &  209,220 &   583,386 & 154,090 \\
\texttt{nom10shuttle}          & 43,500 &  97 &  435,000 &     2931 & 810 \\
\texttt{nom15magic}            & 19,020 & 152 &  209,220 &  1,149,717 & 397,224 \\
\texttt{nom20magic}            & 19,020 & 202 &  209,220 &  1,376,212 & 654,028 \\
\texttt{nom5bike\_day}          & 731   &  65 &    9293 &    61,853 & 16,296 \\
\texttt{nom5bike\_hour}         & 17,379 &  90 &  238,292 &  1,868,205 & 320,679 \\
\texttt{nom5crx}               & 653   &  55 &    8774 &    29,697 & 2162 \\
\texttt{nom5keg}               & 65,554 & 144 & 1,834,566 & 13,262,627 & 42,992 \\
\texttt{nom5shuttle}           & 43,500 &  52 &  435,000 &     1461 & 319 \\\hline
\texttt{ord10bike\_day}        & 731   &  93 &   28,333 &   664,713 & 11,795 \\
\texttt{ord10crx}              & 653   &  79 &   37,005 &  1,547,971 & 2906 \\
\texttt{ord10shuttle}          & 43,500 &  88 & 1,849,216 &    97,357      & 279 \\
\texttt{ord5bike\_day}         & 731   &  58 &   14,929 &    81,277 & 5202 \\
\texttt{ord5bike\_hour}        & 17,379 &  83 &  457,578 &  2,174,964 & 99,691 \\
\texttt{ord5crx}               & 653   &  49 &   19,440 &   139,752 & 973 \\
\texttt{ord5magic}             & 19,020 &  42 &  535,090 &   821,796 & 1267 \\
\texttt{ord5shuttle}           & 43,500 &  43 &  868,894 &     4068 & 119 \\
\texttt{ord6magic}             & 19,020 &  52 &  662,177 &  2,745,877 & 2735 \\
\end{tabular}
\end{sidewaystable}

\medskip

For this batch, we included LinCbO1 (Algorithm~\ref{alg:dgcbo1}) to show how the reuse of attribute 
counters influences the performance.


\begin{sidewaystable}
\setlength{\tabcolsep}{10pt}
\renewcommand{\bfseries}{\fontseries{b}\selectfont} 
\robustify\bfseries             
\newrobustcmd{\BB}{\bfseries}    

\caption{Runtimes in seconds of algorithms generating Duquenne-Guigues basis in batch 2. The symbol $\ast$ means that 
the run could not be completed  due to insufficient memory
\label{tab:big}}
\begin{tabular}{l||
S[table-format=4.3,detect-weight,mode=text]|
S[table-format=4.3,detect-weight,mode=text]|
S[table-format=4.3,detect-weight,mode=text]|
S[table-format=4.3,detect-weight,mode=text]|
S[table-format=4.3,detect-weight,mode=text]|
S[table-format=4.3,detect-weight,mode=text]|
S[table-format=4.3,detect-weight,mode=text]|
S[table-format=4.3,detect-weight,mode=text]|
S[table-format=4.3,detect-weight,mode=text]}

Dataset       & AttInc & NC1 & NC2 & NC3 & {NC$^+$1} & {NC$^+$2} & {NC$^+$3} & {LinCbO} & {LinCbO1} \\
  \hline\hline
\texttt{inter10crx}            &\BB 400.292    &  2084.12   & 17059.5     &  4256.41    &  2097.54   & 16817.5    &   4193.46   &       508.551  & 23842   \\
\texttt{inter10shuttle}        &   $\ast$      & 18038.1    & 21268.1     & 20211.9     & 17664.5    & 21035.4    &  20171.9    &   \BB 15852.9   & 28373.5 \\
\texttt{inter3magic}           &   109.178     &   106.341  &   136.738   &   109.133   &   107.357  &   136.842  &    109.428  &\BB      26.156 & 74.98\\
\texttt{inter4magic}           &   $\ast$      &  4029.95   &  9998.74    &  4241.51    &  4027.48   & 10023      &   4239.26   &\BB   965.353   & 9258.53 \\
\texttt{inter5bike\_day}       &\BB 72.952     &   389.073  &  1409.69    &   680.789   &   383.537  &  1378.89   &    670.109  &        85.591  & 1589.58 \\
\texttt{inter5crx}             &     5.863     &    16.357  &    56.977   &    25.08    &    16.257  &    56.669  &     24.995  &\BB    3.176    & 75.205 \\
\texttt{inter5shuttle}         &   207.323     &   137.211  &   144.747   &   144.125   &   137.596  &   145.491  &    144.957  &\BB   120.003   & 143.4 \\
\texttt{inter6shuttle}         &   253.166     &   164.355  &   181.19    &   177.138   &   164.924  &   182.664  &    178.474  &\BB   133.288   & 181.967 \\\hline
\texttt{nom10bike\_day}        &\BB 4.515      &    42.074  &    33.725   &    71.745   &    31.505  &    24.71 &     52.249  &         7.099  & 26.318 \\
\texttt{nom10crx}              &     1.227     &     3.105  &     5.409   &     7.776   &     2.828  &     4.792  &      6.855  &\BB     0.944   & 6.939 \\
\texttt{nom10magic}            &   486.926     &  1503.38   &   977.612   &  1547.33    &  1322.62   &   790.61   &   1246.06   &\BB   206.797   & 821.269 \\
\texttt{nom10shuttle}          &     1.455     &     1.14   &     1.19    &     1.234   &     1.102  &     1.134  &     1.166   &\BB     0.425   & 0.53 \\
\texttt{nom15magic}            &  3358.44      & 10499.8    &  6442.54    & 14838.1     &  8620.79   &  5060.17   & 11277       &\BB  1509.86    & 5363.77 \\
\texttt{nom20magic}            &  7882.15      & 32600.2    & 16779.1     & 46609.8     & 23129.5    & 10754.4    & 33369.5     &\BB  4437.05    & 17424 \\ 
\texttt{nom5bike\_day}         &     2.58      &    13.064  &    11.32    &    17.572   &    10.855  &     9.383  &    14.517   &\BB     2.219   & 9.251 \\
\texttt{nom5bike\_hour}        & 1893.33       &  8083.01   &  8412.02    &  8402.16    &  7248.4    &  7055.42   &  7163.17    &\BB  1410.11    & 8098.72  \\ 
\texttt{nom5crx}               &     0.406     &     0.623  &     1.054   &     1.061   &     0.592  &     0.983  &     0.988   &\BB     0.193   & 1.110\\
\texttt{nom5keg}               &  $\ast$       &  7707.54   & 16584.8     & 13154.5     &  7564.71   & 16590.3    & 13184.1     &\BB  1936.7     & 15305\\
\texttt{nom5shuttle}           &     0.693     &     0.493  &     0.511   &     0.511   &     0.481  &     0.497  &     0.5     &\BB     0.309   & 0.320\\\hline
\texttt{ord10bike\_day}        &\BB 21.884     &    92.944  &   402.8     &   154.541   &    90.973  &   385.489  &   148.472   &        24.997  & 451\\
\texttt{ord10crx}              &    28.367     &    85.67   &   325.608   &    93.936   &    85.735  &   325.742  &    94.394   &\BB    11.653   & 342.858\\
\texttt{ord10shuttle}          &    51.839     &    40.338  &    42.438   &    41.475   &    40.426  &    42.419  &    41.549   &\BB    34.293   & 40.155\\
\texttt{ord5bike\_day}         &     2.08      &     4.688  &    12.498   &     7.34    &     4.412  &    11.501  &     6.812   &\BB     0.936   & 12.454\\
\texttt{ord5bike\_hour}        &  1107.57      &  1749.29   &  5621.96    &  2304.73    &  1672.93   &  5173.36   &  2169.43    &\BB   321.147   & 5694.64\\
\texttt{ord5crx}               &     1.468     &     2.7    &     6.696   &     3.071   &     2.701  &     6.68 &     3.062   &\BB     0.61  & 6.957\\
\texttt{ord5magic}             &    99.92      &    93.845  &   108.648   &    94.28  &    93.93   &   108.733  &    94.437   &\BB    46.982   & 71.721\\
\texttt{ord5shuttle}           &     1.676     &     1.382  &     1.408   &     1.41  &     1.38   &     1.403  &     1.404   &\BB     1.319   & 1.417\\
\texttt{ord6magic}             &   345.392     &   335.947  &   447.37    &   337.462   &   336.4    &   447.353  &   338.321   &\BB   158.227   & 277.617\\
    
\end{tabular}
\end{sidewaystable}

For most datasets, LinCbO works faster than the other algorithms. For the remaining datasets,
LinCbO is the second best after the attribute incremental approach (see Table~\ref{tab:big}). However,
we encountered limits of the attribute incremental approach as it runs out of available memory
in \green{three} cases (denoted by the symbol $\ast$ in Table~\ref{tab:big}).

\subsection{Evaluation}
\label{sec:discus}

Based on the experimental evaluation in Section~\ref{sec:ee},
we conclude that LinCbO is the fastest algorithm for computation of the Duquenne-Guigues basis. 
In some cases, it is outperformed by the attribute incremental approach. However, the 
attribute incremental approach seems to have enormous memory requirements as it run out of
memory for several datasets.

\bigskip

Originally, we believed that CbO itself can make the computation faster. 
This motivation came from the paper by Outrata \& Vychodil \cite{Outrata2012114},
where CbO is shown to be significantly faster than NextClosure when computing intents (see Table~\ref{tab:nextclosure_cbo}).
\begin{table}[t]
\begin{center}
\begin{tabular}{l|r@{~~~}r@{~~~}r@{~~~}r}
dataset     & mushroom & anonymous web & adult & internet ads \\
size        & 8124 $\times$ 119 & 32,711 $\times$ 296 & 48,842 $\times$ 104 & 3279 $\times$ 1557 \\
fill ratio  & 19.33\,\% & 1.02\,\% & 8.65\,\% & 0.88\,\% \\
\#concepts  & 238,710 & 129,009 & 180,115 & 9192 \\\hline
NextClosure & 53.891 & 243.325 & 134.954 & 114.493 \\
CbO         & 0.508  & 0.238   & 0.302   & 0.332
\end{tabular}
\end{center}
\caption{Runtimes of formal concept enumeration by NextClosure and CbO in seconds for selected datasets (source: \cite{Outrata2012114})\label{tab:nextclosure_cbo}}
\end{table}
The main reason for the speed-up is the fact 
that CbO uses set intersection to efficiently obtain extents during the tree descent.
This feature cannot be exploited for computation of the Duquenne-Guigues basis. 
The CbO itself rarely seems to have a significant effect 
on the runtime -- this was the case for datasets
\texttt{nom10shutle} and \texttt{nom5shutle}.
Sometimes, it lead to worse performance, for example for datasets
\texttt{inter10crx}, \texttt{inter10shuttle}, and \texttt{nom20magic}.

\begin{figure}
\begin{center}
\scalebox{.8}{
\begin{tikzpicture}
\begin{axis}[
 ybar=2pt,
 bar width=4pt,
 width=12cm,height=6cm,
 ymode=log,
 symbolic x coords={
 \texttt{inter10crx},
 \texttt{inter10shuttle},
 \texttt{inter3magic},
 \texttt{inter4magic},
 \texttt{inter5bikeDday},
 \texttt{inter5crx},
 \texttt{inter5shuttle},
 \texttt{inter6shuttle}},
 xtick=data,
 x tick label style={rotate=45,anchor=east},]

\addplot[fill=blue] coordinates {
(\texttt{inter10crx},16817500) 
(\texttt{inter10shuttle},21035400) 
(\texttt{inter3magic},136842)
(\texttt{inter4magic},10023000)
(\texttt{inter5bikeDday},1378890)
(\texttt{inter5crx},56669)
(\texttt{inter5shuttle},145491)
(\texttt{inter6shuttle},182664)
};

\addplot[fill=red] coordinates {
(\texttt{inter10crx},23842000) 
(\texttt{inter10shuttle},28373500) 
(\texttt{inter3magic},74980)
(\texttt{inter4magic},9258530)
(\texttt{inter5bikeDday},1589580)
(\texttt{inter5crx},75205)
(\texttt{inter5shuttle},143400)
(\texttt{inter6shuttle},181967)
};

\addplot[fill=green] coordinates {
(\texttt{inter10crx},508551) 
(\texttt{inter10shuttle},15852900) 
(\texttt{inter3magic},26156)
(\texttt{inter4magic},965353)
(\texttt{inter5bikeDday},85591)
(\texttt{inter5crx},3176)
(\texttt{inter5shuttle},120003)
(\texttt{inter6shuttle},133288)
};

\end{axis}
\end{tikzpicture}}

\scalebox{.8}{
\begin{tikzpicture}
\begin{axis}[
 ybar=2pt,
 bar width=4pt,
 width=12cm,height=6cm,
 ymode=log,
 symbolic x coords={
 \texttt{nom10bike\_day},
 \texttt{nom10crx},
 \texttt{nom10magic},
 \texttt{nom10shuttle},
 \texttt{nom15magic},
 \texttt{nom20magic},
 \texttt{nom5bike\_day},
 \texttt{nom5bike\_hour},
 \texttt{nom5crx},
 \texttt{nom5keg},
 \texttt{nom5shuttle}},
 xtick=data,
 x tick label style={rotate=45,anchor=east},]

\addplot[fill=blue] coordinates {
(\texttt{nom10bike\_day},24710) 
(\texttt{nom10crx},4792) 
(\texttt{nom10magic},790610)
(\texttt{nom10shuttle},1134)
(\texttt{nom15magic},5060170)
(\texttt{nom20magic},10754400)
(\texttt{nom5bike\_day},9383)
(\texttt{nom5bike\_hour},7055420)
(\texttt{nom5crx},983)
(\texttt{nom5keg},16590300)
(\texttt{nom5shuttle},497)
};

\addplot[fill=red] coordinates {
(\texttt{nom10bike\_day},26317) 
(\texttt{nom10crx},6939) 
(\texttt{nom10magic},821269)
(\texttt{nom10shuttle},529)
(\texttt{nom15magic},5363770)
(\texttt{nom20magic},17424000)
(\texttt{nom5bike\_day},9251)
(\texttt{nom5bike\_hour},8098720)
(\texttt{nom5crx},1110)
(\texttt{nom5keg},15305000)
(\texttt{nom5shuttle},320)
};

\addplot[fill=green] coordinates {
(\texttt{nom10bike\_day},7099) 
(\texttt{nom10crx},944) 
(\texttt{nom10magic},206797)
(\texttt{nom10shuttle},425)
(\texttt{nom15magic},1509860)
(\texttt{nom20magic},4437050)
(\texttt{nom5bike\_day},2219)
(\texttt{nom5bike\_hour},1410110)
(\texttt{nom5crx},193)
(\texttt{nom5keg},1936700)
(\texttt{nom5shuttle},309)
};
\end{axis}
\end{tikzpicture}}

\scalebox{.8}{
\begin{tikzpicture}
\begin{axis}[
 ybar=2pt,
 bar width=4pt,
 width=12cm,height=6cm,
 ymode=log,
 symbolic x coords={
 \texttt{ord10bike\_day},
 \texttt{ord10crx},
 \texttt{ord10shuttle},
 \texttt{ord5bike\_day},
 \texttt{ord5bike\_hour},
 \texttt{ord5crx},
 \texttt{ord5magic},
 \texttt{ord5shuttle},
 \texttt{ord6magic}},
 xtick=data,
 x tick label style={rotate=45,anchor=east},]

\addplot[fill=blue] coordinates {
(\texttt{ord10bike\_day},385489)
(\texttt{ord10crx},325742)
(\texttt{ord10shuttle},42419)
(\texttt{ord5bike\_day},11501)
(\texttt{ord5bike\_hour},5173360)
(\texttt{ord5crx},6680)
(\texttt{ord5magic},108733)
(\texttt{ord5shuttle},1403)
(\texttt{ord6magic},447353)
};

\addplot[fill=red] coordinates {
(\texttt{ord10bike\_day},451000)
(\texttt{ord10crx},342858)
(\texttt{ord10shuttle},40155)
(\texttt{ord5bike\_day},12454)
(\texttt{ord5bike\_hour},5694640)
(\texttt{ord5crx},6957)
(\texttt{ord5magic},71721)
(\texttt{ord5shuttle},1417)
(\texttt{ord6magic},277617)
};

\addplot[fill=green] coordinates {
(\texttt{ord10bike\_day},24997)
(\texttt{ord10crx},11653)
(\texttt{ord10shuttle},34293)
(\texttt{ord5bike\_day},936)
(\texttt{ord5bike\_hour},321147)
(\texttt{ord5crx},610)
(\texttt{ord5magic},46982)
(\texttt{ord5shuttle},1319)
(\texttt{ord6magic},158227)
};
\end{axis}
\end{tikzpicture}}

\begin{tikzpicture}
\node[draw,color=black,fill=blue,rectangle,label={0:NC$^+$2}]{~};
\node[draw,color=black,fill=red,rectangle,label={0:LinCbO1}] at (3,0) {~};
\node[draw,color=black,fill=green,rectangle,label={0:LinCbO}] at (6,0) {~};
\end{tikzpicture}
\end{center}
\caption{\label{fig:lincbo} Comparison of NextClosure with LinClosure with an early stop (NC$^+$2, LinCbO1, and LinCbO for datasets in batch~2;
runtimes in milliseconds on a logarithmic scale (values are from Table~\ref{tab:big}).}
\end{figure}
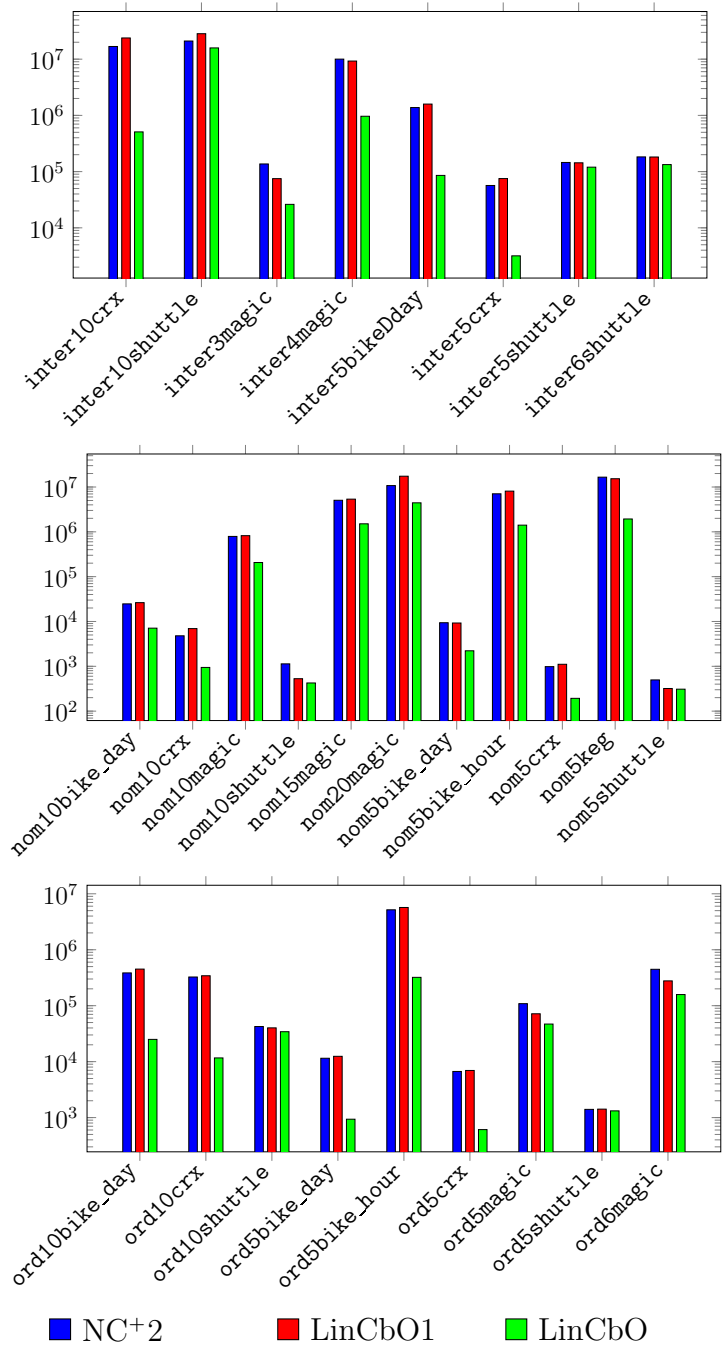

However, the introduction of the reuse of attribute counters 
significantly improves the runtime for most datasets (see Fig.\,\ref{fig:lincbo}).

\section{Conclusions and further research}
\label{sec:conc}

The algorithm LinClosure has been considered to be slow and even worse than the na\"ive closure \cite{wild1995computations,bazhanov2014}. In an experimental evaluation, we have shown that it can perform very fast when it can reuse its attribute counters. The reuse is enabled by using CbO.

As our future research, we want to further develop the present algorithm.
\begin{itemize}
\item
One of the benefits of CbO is that it can be improved to avoid some unnecessary closure computations.
This improvement, called pruning, is in various ways utilized in FCbO \cite{Outrata2012114} and In-Close ver.\,3 and higher
\cite{Andrews2015633,Andrews2017,andrews2018new}. In the case of the Duquenne-Guigues basis, the 
computation of closure is much more time consuming than in the case of intents. Therefore, it seems to be
a good idea to apply pruning techniques in our algorithm. Our preliminary results indicate a possible 20\,\% speed-up.

\item
Generalization of LinClosure is used to compute models in generalized settings, 
like fuzzy attribute implications \cite{DBLP:conf/cla/BelohlavekV06,DBLP:journals/ijgs/BelohlavekV16,DBLP:journals/ijgs/BelohlavekV17} and temporal attribute implications \cite{DBLP:journals/amai/TriskaV18}. We will explore potential uses of LinCbO in these generalizations.

\item 
Algorithms  for enumeration of closed sets can be
extended to handle a background knowledge given as a set of attribute implications or as a constraint closure operator \cite{bevy:bkcknow}.
Adding the background knowledge in the computation of the Duquenne-Guigues basis was investigated by Kriegel \cite{kriegel}.
We will explore this possibility for LinCbO.

\item
The implementation used for experimental evaluation was made to be at a similar level to the Bazhanov and Obiedkov implementations \cite{bazhanov2014}.
We will deliver an optimized implementation of LinCbO, possibly with a pruning technique.
\end{itemize}

\section*{Acknowledgment}

The authors acknowledge support by the grants
\begin{itemize}
\item
IGA~UP~2020 of
Palack\'y University Olomouc, No. IGA\_PrF\_2020\_019,
\item
JG~2019 of 
Palack\'y University Olomouc, No.\,JG\_2019\_008.
\end{itemize}

\section*{References}


\end{document}